\pdfoutput=1
 \documentclass[aps,prl,reprint,twocolumn,showpacs,superscriptaddress]{revtex4-1}  
 \usepackage{auncial}
\usepackage{scrextend}
\usepackage{relsize}
\usepackage{amsmath}
\usepackage{amssymb}
\usepackage{epsfig}
\usepackage{graphicx}
\usepackage{hyperref}
\usepackage{dcolumn}   % needed for some tables
\usepackage{tabu}
\usepackage{boldline}
\usepackage{slashed}
\usepackage{multirow}
\usepackage{color}
\usepackage[normal]{subfigure}
\usepackage{rotating}
\usepackage[margin=0.9in,a4paper]{geometry}
\usepackage[table]{xcolor}
\usepackage{enumitem}
\usepackage[utf8]{inputenc}
\usepackage{colortbl}
\usepackage{array,multirow}
\definecolor{nicered}{rgb}{0.7,0.1,0.1}
\definecolor{nicegreen}{rgb}{0.1,0.5,0.1}
\definecolor{red}{rgb}{1.0, 0, 0}

\def\gsim{\raise0.3ex\hbox{$\;>$\kern-0.75em\raise-1.1ex\hbox{$\sim\;$}}}
\def\lsim{\raise0.3ex\hbox{$\;<$\kern-0.75em\raise-1.1ex\hbox{$\sim\;$}}}
\def\mb[#1]{\mathbf{#1}}
\renewcommand{\bar}{\overline}

\definecolor{LightCyan}{rgb}{0.88,1,1}
\definecolor{piggypink}{rgb}{0.99, 0.87, 0.9}
\definecolor{applegreen}{rgb}{0.55, 0.71, 0.0}
\definecolor{darkpastelgreen}{rgb}{0.01, 0.75, 0.24}
\definecolor{green-yellow}{rgb}{0.68, 1.0, 0.18}

\newcommand{\beq}{\begin{equation}}
\newcommand{\eeq}{\end{equation}}
\newcommand{\beqa}{\begin{eqnarray}}
\newcommand{\eeqa}{\end{eqnarray}}
\newcommand{\Sec}[1]{\section{#1}}

\newcommand{\HD}{\mathcal{H}}
\newcommand{\Mp}{M_{\rm Pl}}

\def\beqn{\begin{eqnarray}} 
\def\eeqn{\end{eqnarray}} 
\def\be{\begin{equation}}
\def\ee{\end{equation}}
\def\nn{\nonumber}

\begin{document}
% ----------------- preprint numbers ------------------
% \begin{frontmatter}

% ------------- Title and authors ---------------------

\title{Relaxed Inflation}

\author{Walter Tangarife}
\email{waltert@post.tau.ac.il}
\affiliation{\normalsize \it 
Raymond and Beverly Sackler School of Physics and Astronomy, \\
 Tel-Aviv University, Tel-Aviv 69978, Israel}
\author{Kohsaku Tobioka}
\email{kohsakut@post.tau.ac.il}
\affiliation{\normalsize\it 
Raymond and Beverly Sackler School of Physics and Astronomy, \\
 Tel-Aviv University, Tel-Aviv 69978, Israel}
 \affiliation{\normalsize\it
 Department of Particle Physics and Astrophysics, \\
Weizmann Institute of Science, Rehovot 76100, Israel}
\author{Lorenzo Ubaldi}
\email{ubaldi.physics@gmail.com}
\affiliation{\normalsize\it 
Raymond and Beverly Sackler School of Physics and Astronomy, \\
 Tel-Aviv University, Tel-Aviv 69978, Israel}
 \author{Tomer Volansky}
\email{tomerv@post.tau.ac.il}
\affiliation{\normalsize\it 
Raymond and Beverly Sackler School of Physics and Astronomy, \\
 Tel-Aviv University, Tel-Aviv 69978, Israel}

% ------------------------------------------------------
\begin{abstract}
  \noindent
We present an effective model where the inflaton is a relaxion that scans the Higgs mass and sets it at the weak scale. 
The dynamics consist of a long epoch in which inflation is due to the shallow slope of the potential,
followed by a few number of e-folds where slow-roll is maintained thanks to dissipation via non-perturbative gauge-boson
production. The same gauge bosons give rise to a strong electric field that triggers the production of electron-positron
pairs via the Schwinger mechanism. The subsequent thermalization of these particles provides a novel mechanism of reheating.
The relaxation of the Higgs mass occurs after reheating, when the inflaton/relaxion stops on a local minimum of
the potential.
We argue that this scenario may evade phenomenological and astrophysical bounds while allowing for the  cutoff of the effective model to be close to the Planck scale. This framework provides an intriguing 
connection between inflation and the hierarchy problem.
 
\end{abstract}

%%%%%%%%%%  \pacs{14.80.Va, 14.65.Jk}
%%%%%%%%%%  \keywords{Strong CP problem, Axions}
% \end{keyword}
%%  \PACS 
% \end{frontmatter}

 \maketitle

 \Sec{Introduction}  

The mass of the Higgs boson, $m_h$, is sixteen orders of magnitude
smaller than the Planck mass. This poses a puzzle, which goes under the name of the naturalness problem.
In the Standard Model (SM) of particle physics, we expect large quantum corrections
that would raise $m_h$ roughly up to the Planck scale. One way to avoid such corrections
is to impose additional symmetries to protect $m_h$, and keep it naturally small. 
Supersymmetry \cite{Martin:1997ns} is the most studied extension in this direction and, like  
 most other solutions, predicts the presence  of new physics
at around the TeV scale that can potentially be accessible 
 at the Large Hadron Collider.  

Another direction in addressing the problem of naturalness has been put forward in Ref.~\cite{Graham:2015cka}.
The smallness of $m_h$ could result from the cosmological evolution of another scalar field, the relaxion, that
couples to the Higgs, scans its mass and eventually sets it to the observed value. This solution is based on dynamics
rather than symmetry~\footnote{This is true modulo the fact that it relies on an argument of technical naturalness, on which 
we elaborate further in the next section.}, and provides an intriguing connection between naturalness and cosmology.
The model is described by an effective Lagrangian valid up to a cutoff scale $\Lambda$, and the success in addressing the small Higgs mass is measured by how high $\Lambda$ is compared to $m_h$, once the constraints from the dynamics
are taken into account. In the original proposal, the highest $\Lambda$ is of order $10^8$ GeV, and can be achieved
in a scenario where the relaxation dynamics take place during inflation. Various features of this class of models have been explored in Refs.~\cite{Espinosa:2015eda, Hardy:2015laa, Jaeckel:2015txa, Gupta:2015uea, Batell:2015fma, Matsedonskyi:2015xta, Marzola:2015dia, Choi:2015fiu, Kaplan:2015fuy, DiChiara:2015euo, Ibanez:2015fcv, Hebecker:2015zss, Fonseca:2016eoo, Fowlie:2016jlx, Evans:2016htp, Huang:2016dhp, Kobayashi:2016bue, Hook:2016mqo, Higaki:2016cqb, Choi:2016luu, Flacke:2016szy, McAllister:2016vzi, Choi:2016kke, Lalak:2016mbv, You:2017kah, Evans:2017bjs}.
In this letter, we take the idea of Ref.~\cite{Graham:2015cka} a step further by promoting the relaxion to an inflaton.
The advantages of doing so are that (i) the model is more minimal, as it does not have to rely on an
unspecified inflaton sector, and (ii) it evades numerous constraints, allowing the cutoff to lie close to the Planck scale.

In the rest of the paper we describe the model, the dynamics of inflation, a novel reheating mechanism, and the relaxation of the 
electroweak (EW) scale, which happens after reheating. The interested reader can find more details in a longer
companion paper \cite{longpaper}.

%%%%%%%%%%%%%%%%%%
\Sec{The model}

We consider the effective Lagrangian
\begin{align}
&\mathcal{L} =   -\frac{1}{2} \partial_\mu \phi \partial^\mu \phi  -\frac{1}{4} F_{\mu\nu}F^{\mu\nu} - c_\gamma \frac{\phi}{4f} F_{\mu\nu} \tilde F^{\mu\nu} \nonumber \\
&\quad\quad - (g_h m \phi - \Lambda^2) \HD^\dagger \HD - \lambda (\HD^\dagger \HD)^2  \nonumber \\
&\quad\quad - V_{\rm roll} (\phi) - V_{\rm wig}(\phi) - V_0  \, , \label{eq:Lagrangian} \\
& V_{\rm roll} (\phi) =   m \Lambda^2 \phi  \, , \label{Vrolldef} 
 \quad V_{\rm wig}(\phi) =  \Lambda_{\rm wig}^4 \cos \frac{\phi}{f} \, , %\label{Vwigdef}
\end{align}
defined in a Friedmann-Robertson-Walker (FRW) metric, $ds^2 = - dt^2 + a^2(t) d\vec x^2$. Here, $\phi$ is the relaxion/inflaton,
$\HD$ the Higgs doublet, $F_{\mu\nu}$ the field strength of an Abelian gauge field, $\tilde F_{\mu\nu}$ its dual.   $f$ is the scale of spontaneous breaking of a global $U(1)$, 
of which $\phi$ is the Goldstone boson.
$g_h$ is a dimensionless coupling of order one, $c_{\gamma}$ is model dependent and can span a large range of values.
$\Lambda$ is the bare Higgs mass and the cutoff of the effective theory.

The relaxion potential has three terms: $V(\phi) =V_{\rm roll} (\phi) + V_{\rm wig} (\phi) +V_0$. 
The first is responsible for the rolling, and is linear in $\phi$ (we neglect higher powers, 
which would come with correspondingly higher powers of the small mass parameter $m$). The second is responsible for the periodic potential (``wiggles"),
which grows proportionally to the Higgs vacuum expectation value (VEV), $v$, as
$\Lambda_{\rm wig}^4 \sim (yv)^n M^{4-n}$. Here, $y$ is a Yukawa coupling and $M$ is a mass scale smaller than $4\pi v$. 
Note that for $n$ odd the wiggles are present only when $\HD$ has a nonzero VEV, while for $n$ even
they are present also in the unbroken EW phase~\cite{Espinosa:2015eda, Gupta:2015uea}. In what follows,
we concentrate, for simplicity, on the QCD-like case, $n=1$.
The third term, $V_0$, is a constant that we choose to set $V(\phi)=0$ at the local minimum where we obtain the correct EW scale,
\be
\langle |\HD|\rangle \equiv v \simeq \frac{m_W}{\sqrt{\lambda}} \, , \, \,  \langle \phi \rangle \equiv \phi_{\rm EW} \simeq
\frac{1}{g_h m} (\Lambda^2 - m_W^2) \, .
\ee
One finds
\be \label{V0def}
V_0 = -\frac{\Lambda^4}{g_h}+ \frac{m_W^2 \Lambda^2}{g_h} + \frac{m_W^4}{4 \lambda} \, . 
\ee 
Choosing this $V_0$ corresponds to tuning the cosmological constant. This is crucial, as it determines the dynamics
of the field and ensures the exit of inflation before the relaxion settles into the EW vacuum.

An important ingredient is that the mass parameter $m$, which controls the slope of the rolling potential, is
tiny. This is technically natural, since in the limit $m \to 0$ the Lagrangian recovers the discrete shift symmetry
$\phi \to \phi + 2\pi f$.
The scales in the model have the following hierarchic structure
\be 
m \ll \Lambda_{\rm wig} < 4\pi m_W \ll \Lambda <  f < \Mp \, ,
\ee  
where $\Mp$ is the reduced Planck mass.

 The relaxion is coupled to an Abelian gauge field, $A_\mu$. The time-dependent relaxion background eventually leads
 to an exponential production of long-wavelength modes of the gauge bosons. This has two important consequences: 
 (i) it provides a new source of dissipation for inflation, also maintaining slow-roll during the final period of the relaxation, which necessarily occurs after the end of inflation;
 (ii) it allows for a novel reheating mechanism that proceeds as follows. The gauge bosons form a strong coherent electric
 field that produces electron-positron pairs via the Schwinger effect. These particles quickly thermalize and reheat the universe.
 We discuss these steps in more detail in the following sections.
 Note that since the relaxion is very light and weakly coupled, reheating mechanisms via perturbative decays are not effective
 in this framework.

%%%%%%%%%%%%%%%%%%%%%%%%%%%%%%%%%%%%%%%%%%%
\Sec{Inflationary dynamics} 
\begin{figure} [t]
\begin{center}
 \includegraphics[width=0.48\textwidth]{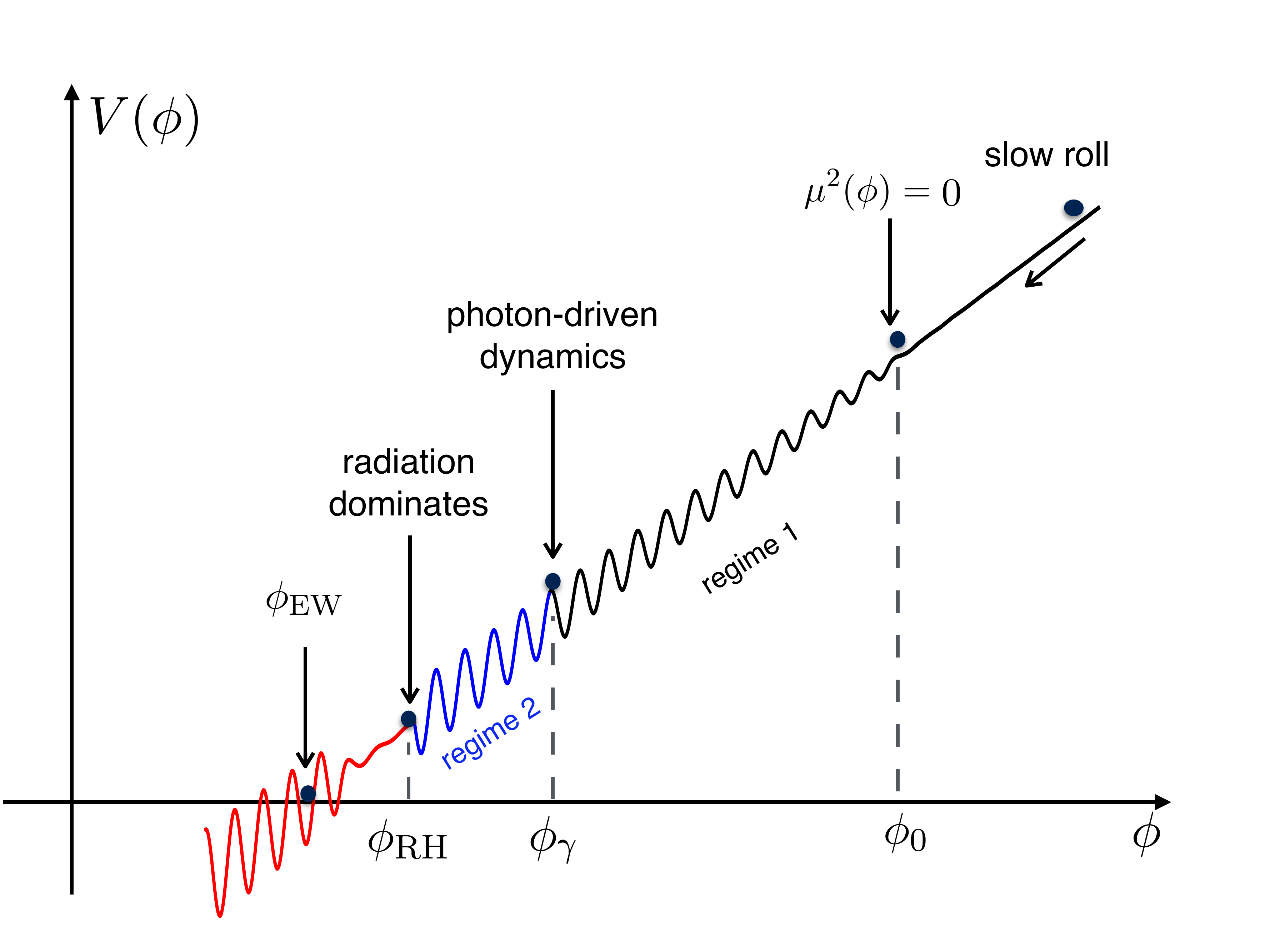}
 \caption{Sketch of evolution stages. The first regime (black) corresponds to inflation via standard slow-roll on a shallow slope. In the second regime (blue), the (dark) photons are responsible for the dissipation. Relaxation occurs in the last stage, after reheating (red). } \label{Fig:cartoon}
 \end{center}
\end{figure}

The equation of motion (EOM) of the inflaton/relaxion in the FRW metric is
\be
 \ddot\phi + 3 H \dot\phi + V'(\phi) = \frac{c_\gamma}{f} \langle \vec E \cdot \vec B \rangle \, , \label{phiEOM} \\
\ee
where $\vec E$ and $\vec B$ are the electric and magnetic fields associated with the gauge field.
Throughout this paper, an overdot denotes a derivative with respect to cosmic time, $t$.
 The energy density
of the universe is dominated by the relaxion, that drives inflation, so the Hubble parameter is $H = \frac{\sqrt{V(\phi)}}{\sqrt{3} \Mp}$.
In Eq.~\eqref{phiEOM}, we have neglected the term $g_h m \langle |\HD|^2 \rangle$, because it is always negligible compared
to $V'(\phi)$. 

The dynamics can be described in three different stages, illustrated in Fig.~\ref{Fig:cartoon}. 
The rolling starts at large values of the field, $\phi > \phi_0 \equiv \frac{\Lambda^2}{g_h m}$,
where $\HD$ has no VEV, and consequently there is no periodic potential.
With our conventions, $\phi$ moves from right to left.
 In the first stage, the EOM is 
\be \label{EOMreg1}
3 H \dot\phi + V'(\phi) = 0 \,
\ee
to a very good approximation, and the relaxion rolls slowly due to the shallow linear slope. The speed,
$|\dot\phi| = \frac{V'(\phi)}{3H}$, slowly increases as $H$ decreases going down the potential, but
stays small enough so that $\langle \vec E \cdot \vec B \rangle$  is negligible at this stage [see Eq.~\eqref{edotb} below].
This regime involves trans-Planckian field excursions, 
lasts for a very large number of e-folds, $N > 10^{30}$, and continues into the broken
EW phase, $\phi < \phi_0$. Eventually the speed grows large enough that the gauge-boson production becomes
the dominant source of friction. To understand how that happens, we turn our attention to the EOM of the massless 
gauge field.

After expanding $A_\mu$ in Fourier modes, the EOM for the two polarizations reads~\cite{Anber:2009ua}
\be
\label{AEOM}
\frac{\partial^{2}A_\pm^{\vec k}(\tau)}{\partial\tau^{2}}+\left[k^{2}\pm 2\,k\,\frac{\xi}{\tau} \right]A_\pm^{\vec k}(\tau)=0 \, ,
\ee
where $\tau$ is the conformal time, $d\tau = dt / a$, and we have defined 
\begin{equation} \label{xidef}
\xi\equiv c_\gamma\frac{\dot\phi}{2\,f\,H}\,\,.
\end{equation}
Note that $\tau$ and $\xi$ are both negative. Eq. \eqref{AEOM} implies that low-momentum (long-wavelength) 
modes of $A_-^{\vec k}(\tau)$, satisfying $k -2 \frac{\xi}{\tau} < 0$, experience tachyonic instability and grow exponentially. 
The solution can be written approximately as
\begin{equation}
\label{approx}
A_-^{\vec k}(\tau)\simeq 
\frac{1}{\sqrt{2\,k}}\left(\frac{ -k \tau}{2\,|\xi|}\right)^{1/4}e^{\pi\,|\xi|-2\,\sqrt{- 2|\xi| \,k \, \tau}} \,  ,
\end{equation}
for $|\xi| > 1$, and we can use it to compute
\begin{align}
\langle  \vec E\cdot \vec B \rangle &
  \simeq 2.4\times 10^{-4} \frac{H^4}{|\xi|^4} e^{2\pi |\xi|}  , \label{edotb}
 \\
\langle  \vec E^2 \rangle &
\simeq 10^{-4}  \frac{H^4}{|\xi|^3} e^{2\pi |\xi|} \, , \quad 
\langle  \vec B^2 \rangle 
\simeq 10^{-4}  \frac{H^4}{|\xi|^5} e^{2\pi |\xi|}  . \label{rhog}
\end{align}
Once $|\dot\phi|$, and hence $|\xi|$, grow large enough, we smoothly switch from Eq.~\eqref{EOMreg1} 
to the EOM
\be \label{EOMreg2}
V'(\phi) = \frac{c_\gamma}{f} \langle \vec E \cdot \vec B \rangle \, ,
\ee
where the dissipation is due to gauge-boson production. The solution now is 
\begin{align}
|\dot\phi| = 2|\xi| \frac{H f}{c_\gamma}
 \simeq  \frac{H f}{\pi c_\gamma} \ln \left[\frac{|\xi|^4}{2.4 \times10^{-4}H^4}  \frac{f V'(\phi)}{c_\gamma}   \right] \, .
\end{align}
In this regime, $|\xi| \sim 20$ is roughly constant (only varies logarithmically), and $|\dot\phi|$ decreases with
the decreasing $H$. 
The energy density of the gauge bosons, 
$\rho_\gamma = \frac{1}{2} \langle \vec E^2 + \vec B^2 \rangle$, is roughly constant, 
and using Eq.~\eqref{EOMreg2} we have the relation
$\rho_\gamma \simeq \frac{|\xi|}{c_\gamma} f V'(\phi)$. 
One can show that the slow-roll conditions are now satisfied as long as~\cite{longpaper}
\be \label{fupper}
\frac{f}{c_\gamma} < \frac{\Mp}{|\xi|} \, .
\ee 

When the potential $V(\phi)$ attains a value smaller than $\rho_\gamma$, the energy density 
is no longer dominated by the inflaton and we exit inflation. The following evolution is still 
described by Eq.~\eqref{EOMreg2}, the relaxion keeps slowing down and its kinetic energy
remains smaller than $\Lambda_{\rm wig}^4$. This implies that as the periodic wiggly potential becomes sufficiently large to balance the linear slope, the field stops. Specifically, this condition reads
\be
m \Lambda^2 \sim \frac{\Lambda_{\rm wig}^4}{f} \, .
\ee
This must happen when $\phi = \phi_{\rm EW}$.
By taking $m$ very small, we can achieve a very large $\Lambda$, the only bound being 
$\Lambda < f < \Mp$. Therefore, with a large $f$, we can have a cutoff $\Lambda$
close to the Planck scale. In the original proposal~\cite{Graham:2015cka}, where $\phi$ was 
not the inflaton, $\Lambda$ was mainly constrained by the requirements that the inflaton
dominate the energy density and that the classical motion of the relaxion dominate over
its quantum fluctuations. Neither requirement is necessary in our framework, which allows for
a significantly larger cutoff scale.   Further details of the phenomenological and astrophysical constraints and the corresponding viable parameter space can be found in~\cite{longpaper}.

The picture of this section seems to describe a successful model of inflation that 
relaxes the EW scale. However, there are some subtle complications related to the gauge-boson
production that we have to face. They are the subject of the next sections, where we show how they 
lead to a novel mechanism of reheating.

%%%%%%%%%%%%%%%%%%%%%%%%%%%%%%%%%%%%%%%%%%%%%%%
\Sec{Schwinger effect and reheating}

The produced gauge bosons have a comoving wavelength comparable to the size of the comoving
horizon, $(aH)^{-1}$, and an exponentially large occupation number. They form a coherent collection
that describes a classical electromagnetic field with the electric field dominating the energy density
[see Eq.~\eqref{rhog}], and approximately constant within the horizon.
If $A_\mu$ is the SM photon, the strong electric field allows for electron-positron pair production, via
the Schwinger mechanism, with a rate per unit time per unit volume~\cite{Heisenberg:1935qt, Schwinger:1951nm}
\begin{align}  \label{eq:schwingerrate}
\frac{\Gamma_{e^+e^-}}{V}
= \frac{(e |\vec{E}|)^2}{4\pi^3}\exp \left(\frac{-\pi m_e^2}{e |\vec{E}|}\right) \, .
\end{align}
The production is efficient when $e |\vec E| \geq \pi m_e^2$. 
The $e^+$ and $e^-$ quickly thermalize the system via annihilations, $e^+ e^- \to \gamma \gamma$,
 and inverse Compton scatterings on the long-wavelength photons, $e \gamma \to e \gamma$. The rate of these
 processes is much faster than the rate at which  the electric field is produced by the relaxion. The latter process
 becomes even less efficient after thermalization, because the photon gets a thermal mass which strongly suppresses the tachyonic instability~\cite{Hook:2016mqo, Choi:2016kke}. 
As a result the electric field does not grow larger than $\sim \pi m_e^2/e$ and, consequently, the energy density
 transferred to the $e^+ e^-$ is of order $m_e^4$. This translates into a reheat temperature 
 $T_{\rm RH} \sim m_e$, that is below the Big Bang Nucleosynthesis (BBN) temperature. 
 Moreover, the limited growth of the electric field implies that
 the term $\frac{c_\gamma}{f} \langle \vec E \cdot \vec B \rangle$ remains negligible in Eq.~\eqref{phiEOM}, and we 
 never enter the regime in which the photon dissipation dominates. The kinetic energy of the inflaton then increases above 
 $\Lambda_{\rm wig}^4$ and we overshoot the minimum at $\phi_{\rm EW}$, thus failing to relax the EW scale.
 Fortunately, there is a simple fix to these problems, as we discuss in the next section.

%%%%%%%%%%%%%%%%%%%%%%%%%%%%%%%%%%%%%%%%%%%%%%%%
\Sec{Dynamics with a dark photon}

Instead of SM photons, let us consider the production of dark photons. We modify the first line of Eq.~\eqref{eq:Lagrangian} to 
\begin{align}
\mathcal{L} \, = & \, -\frac{1}{2} \partial_\mu \phi \partial^\mu \phi  -\frac{1}{4} F_{\mu\nu}F^{\mu\nu} -\frac{1}{4} F_{D,\mu\nu}F_D^{\mu\nu}  \nn \\
& -\frac{\kappa}{2}F_{\mu\nu} F^{\mu\nu}_D - c_{\gamma_D} \frac{\phi}{4f} F_{D,\mu\nu} \tilde F_D^{\mu\nu}
   +e A_\mu \bar \psi_e \gamma^\mu \psi_e  \, . \label{Eq:Lag_dark}
\end{align}
The subscript $D$ denotes a dark photon, that kinetically mixes with the SM photon. 
We assume that $\phi$ only couples to $A^\mu_D$ and there is no light content in the dark sector other than
the dark photon. The field redefinition $A^\mu \to A^\mu - \kappa A_D^\mu$ removes the kinetic mixing
and introduces a coupling $\kappa e$ between the dark photon and the SM electrons, $\psi_e$.

The relaxion dynamics proceeds in the same way as described above, the only difference being that now we produce
dark electric and magnetic fields. The important point is that the Schwinger rate changes to
\begin{align}  \label{eq:schwingerrateDark}
\frac{\Gamma_{e^+e^-}}{V}
= \frac{(\kappa e |\vec{E_D}|)^2}{4\pi^3}\exp \left(\frac{-\pi m_e^2}{\kappa e |\vec{E_D}|}\right) . 
\end{align} 
This implies that $|\vec{E}_D|$ has to grow larger than $\pi m_e^2 / (\kappa e)$ for the $e^+ e^-$ production 
to occur. The highest value achievable by the electric field, before it saturates the EOM of Eq.~\eqref{EOMreg2},
is
\begin{align}
|\vec{E}_D^{\rm max}|^2 \sim  \rho_{\gamma_D} \simeq  \frac{|\xi|}{c_{\gamma_D}} f V'(\phi) \sim  \frac{|\xi|}{c_{\gamma_D}} \Lambda_{\rm wig}^4 \ .
\end{align}
 This imposes a lower bound on $\kappa e$ to allow for the Schwinger pair creation. 
Meanwhile, to avoid thermal suppression of the tachyonic production of the dark photon, we require that its mean free path
through the hot plasma of $e^+ e^-$ be longer than a Hubble radius. This sets an upper bound on $\kappa e$
and guarantees that $A^\mu_D$ does not get a thermal mass. These two bounds restrict $\kappa e$ to the window
\begin{align}
\frac{m_e^2}{\Lambda_{\rm wig}^2}  \left( \frac{c_{\gamma_D}}{|\xi |} \right)^{1/2}
\lesssim \kappa e \lesssim \left(\frac{\Lambda_{\rm wig}}{\alpha \Mp}  \right)^{1/2} \, ,
\end{align}
which implies a lower bound on $\Lambda_{\rm wig}$:
\be \label{lowerLbr}
\Lambda_{\rm wig} > \left(\alpha \frac{c_{\gamma_D}}{|\xi |} \Mp m_e^4 \right)^{1/5}.
\ee 
Here $\alpha = e^2/(4\pi)$.

At the beginning of the Schwinger production, the energy density of $e^+ e^-$ is of order $m_e^4$, while that of the dark electric 
field is $(\kappa e)^{-2}$ larger. As $|\vec E_D|$ keeps growing
to its maximum value, it shares its energy with the $e^+ e^-$ pairs by accelerating them classically.
At the end of the process we have $\rho_{e^+ e^-} \sim \rho_{\gamma_D}$. 
This is the energy density available for reheating the visible sector. We can thus achieve a reheat temperature
$T_{\rm RH} \sim \left( \frac{|\xi|}{c_{\gamma_D}} \right)^{1/4} \Lambda_{\rm wig}$, safely above BBN.
Due to the lack of thermal suppressions, the EOM of the relaxion is still described by Eq.~\eqref{EOMreg2} 
after reheating. Therefore, the continued friction provided by unsuppressed dark photon production crucially slows down
the motion of $\phi$ and allows it to settle at the EW vacuum. 

Given the small values of $\kappa e$ under
consideration, the dark photons never reach equilibrium with the visible sector and remain cold 
(they have very low momentum) throughout the
thermal history of the universe. In this way, cosmological bounds on relativistic species are evaded. 
What we have is a cold dark electric field, whose energy density, $\rho_{\gamma_D}$, redshifts like radiation and remains comparable
to that of the visible sector until the time of matter - radiation equality. After that point the universe enters the  
matter dominated era, and $\rho_{\gamma_D}$ eventually becomes a negligible component of the energy density budget.
   
There is one more constraint we need to impose on the model. If the gauge-boson production regime lasts too long, 
we overproduce curvature perturbations, non-Gaussianities and primordial black holes~\cite{Anber:2009ua, Barnaby:2011vw,Linde:2012bt, Garcia-Bellido:2016dkw}. To comply with the corresponding CMB bounds we require that we enter this regime only
in the last five e-folds of inflation. This sets a lower bound on $f/c_{\gamma_D}$, and together with the condition
of Eq.~\eqref{fupper} restricts it to the window
\be
0.2 \frac{\Mp}{|\xi|} \lesssim \frac{f}{c_{\gamma_D}} < \frac{\Mp}{|\xi|} \, .
\ee
The above fixes $f$ to be of order $f \simeq c_\gamma M_{\rm Pl}/|\xi|$.
For values of $c_{\gamma_D}$ of order one or larger, $f$ can be close to the Planck scale.
This, in turn, allows for a large cutoff $\Lambda$.

%%%%%%%%%%%%%%%%%%%%%%%%%%%%%%%%%%%%%%%%%%%%%%%
\Sec{Summary} 

We have presented a model where the relaxion, coupled to the Higgs and to a dark photon,
drives inflation and relaxes the EW scale after reheating. Inflation proceeds in two stages. 
In the first, which lasts very long, the relaxion slowly rolls down a shallow slope. In the second, which takes place
only in the last five e-folds, the slow-roll is maintained thanks to dark photon production, that provides dissipation.
The dark photons, kinetically mixed with the SM photons, form a very large dark electric field which
 produces SM $e^+ e^-$ pairs via the Schwinger effect. The $e^+ e^-$ thermalize 
the visible sector to a temperature above BBN. After the reheating process, the relaxion keeps rolling
and slowing down, due to the continued dark photon dissipation, until it stops on the periodic potential and relaxes
the EW scale.

The mechanism realizes a low-scale model of inflation (with $H \sim \Lambda^2_{\rm wig}/\Mp < m^2_W/\Mp$
in the final observable e-folds) that fully addresses at the same time the hierarchy problem
of the Standard Model. Additional details are presented in a companion paper~\cite{longpaper}.
  The associated CMB signatures deserve further detailed studies, as does the  
novel reheating mechanism.  Both will be presented in a future publication.

\section*{Acknowledgments}
We would like to thank Tim Cohen, Erik Kuflik, Josh Ruderman, and Yotam Soreq for collaboration at the embryonic stages
of this work. We benefited from a multitude of discussions with P. Agrawal, B. Batell, C. Csaki, P. Draper, S. Enomoto, W. Fischler, R. Flauger, P. Fox,
R. Harnik, A. Hook, K. Howe,  S. Ipek, J. Kearney, H. Kim, G. Marques-Tavares, L. McAllister, M.~McCullough, S. Nussinov, S. Paban, E. Pajer, M.~Peskin, G. Perez, D. Redigolo, A. Romano, R. Sato, L. Sorbo, and M. Takimoto.
This work is supported in part by the I-CORE Program of the Planning Budgeting Committee and the Israel Science Foundation (grant No. 1937/12), by the European Research Council (ERC) under the EU Horizon 2020 Programme (ERC- CoG-2015 - Proposal n. 682676 LDMThExp) and by the German-Israeli Foundation (grant No. I-1283- 303.7/2014).
The work of LU was performed in part at the Aspen Center for Physics, which is supported by National Science Foundation grant PHY-1066293, and was partially supported by a grant from the Simons Foundation.

 \bibliographystyle{apsrev4-1}
  \bibliography{Relax}

\end{document}